\documentclass[twocolumn,showpacs]{revtex4}
\usepackage{graphicx}%
\usepackage{amsmath}
\usepackage[dvips]{epsfig}      % to include PostScript Figures
\usepackage{subfigure}
\usepackage{array}
\usepackage{color}
\usepackage{ulem}
\usepackage{amssymb}
\usepackage{natbib}

\makeatletter
\def\btt#1{\texttt{\@backslashchar#1}}%
\DeclareRobustCommand\bblash{\btt{\@backslashchar}}%
\makeatother

\begin{document}

\title[Short Title]{Single Boron Atom Anchored on graphitic carbon nitride nanosheet (B/g-C$_2$N) as a photocatalyst for Nitrogen fixation: A First-Principles Study}
\author{Hao-Ran Zhu, Shi-Hao Wei and Da-Yin Hua}
\affiliation{Department of Microelectronic Science and Engineering, School of Physical Science and Technology, Ningbo University, Ningbo, 315211, P.R. China}
\email{weishihao@nbu.edu.cn}%

\date{\today}

\begin{abstract}
Photocatalytic nitrogen reduction is the promising way for ammonia production, the question now is that the search of highly active and low active catalysts. Based on the first-principles calculation, single boron atom is anchored on the g-C$_2$N to form B/g-C$_2$N, the results show that B/g-C$_2$N can serve as a potential photocatalyst for N$_2$ fixation. The introduction of B atom to g-C$_2$N, the energy gap will reduce from 2.45 eV to 1.21 eV, and also show strong absorption in the visible light region. In addition, N$_2$ can be efficiently reduced on B/g-C$_2$N through the enzymatic mechanism  with low onset potential of 0.07 V and rate-determing barrier of 0.50 eV. The ``acceptance-donation'' interaction between B/g-C$_2$N and N$_2$ plays a key role to active N$_2$, the BN$_2$ moiety of B/g-C$_2$N acts as active and transportation center. And the activity originates from the strong interaction between 1$\pi$1$\pi$* orbitals of N$_2$ and molecular orbitals of B/g-C$_2$N, the ionization of 1$\pi$ orbital and the filling of 1$\pi$* orbital can increase the N$\equiv$N bond length greatly, making the activation of N$_2$. Overall, this work demonstrates B/g-C$_2$N is a promising photocatalyst for N$_2$ fixation.

\end{abstract}

%\pacs{36.40.Mr, 31.15.Ew}

\maketitle

\section{INTRODUCTION}
%\label{sec1}
Ammonia (NH$_3$) is not only an important chemical in agriculture and industry fields but also a potential energy storage intermediate due to its high energy density, safe and environment ecofriendly\cite{1,2,3,4}. Currently, the industrial NH$_3$ production mainly relies on the Haber-Bosch process with Fe- or Ru-based catalysts, which involves harsh conditions (typically 300-500 $^{\circ}$C and 200-300 atm), yielding roughly 500 million tons per year, accompanied by the heavy energy consumption, necessary H$_2$ feedstock from fossil fuel and a large number of greenhouse gases emission\cite{3,5,6}. Thus, it is of great significance to develop a green and sustainable strategy for NH$_3$ production\cite{7}. The proton-assisted photocatalytic or electrocatalytic N$_2$ reduction reaction (NRR), ideally under ambient conditions using renewable solar and wind energy, has been proposed as a promising alternative for nitrogen fixation and ammonia production, which stem from N$_2$ biological fixation with nitrogenase enzymes in bacteria that perform nitrogen fixation at room temperature and atmospheric pressure\cite{3,6,8,9,10,11}. For instance, photo(electro)catalytic NRR can be directly driven the sunlight\cite{4,6}. Thus, these two strategies are highly promising for achieving clean, carbon-free and sustainable NH$_3$ production from N$_2$.

Up to now, numerous catalysts have been fabricated for N$_2$ fixation, such as Ru, graphene, MXene, metal oxides, and so on\cite{12,13,14,15,16,17}. Although NRR catalysts based on noble metals (e.g., Au$_{27,28}$, Ru$_{29}$, Rh$_{30}$) show favorable activity, but the scarcity and high cost hinder the widespread application. Compared to metal based catalysts, metal-free catalysts can possess the fact that excellent activity, long durability, and intrinsic advantages of low cost and environmental friendliness\cite{14,16,18,19,20,21,22,23,24,25,26,27,28,29}. Recently, boron-doped graphene shows NH$_3$ production rate can reach up to 9.8 ¦Ìg$\ast$hr$^{-1}$$\ast$cm$^{-2}$ with a high FE of 10.8\% at 0.5 V versus RHE) in aqueous solutions at ambient conditions\cite{14}. B$_4$C nanosheet achieves a high NH$_3$ yield of 26.57 ¦Ìg$\ast$h$^{-1}$ mg$^{-1}$ and a high FE of 15.95\% at a potential of -0.75 V in 0.1 M HCl\cite{16}. The development of efficient metal-free catalysts for NRR of great economic interest and scientific importance. In 2015, a novel 2D layered crystal with uniformly distributed holes, graphitic carbon nitride named g-C$_2$N was successfully prepared by a simple bottom-up wet chemical method\cite{30}. The native porous structureof g-C$_2$N makes it the excellent substrate to support single-atom catalysts (SACs). At present, the N-containing big holes could anchor TM atoms tightly to carry out a series of electro-catalytic reaction like hydrogen evolution reaction (HER)\cite{31,32}, oxygen evolution reaction (OER)\cite{32,33} and N$_2$ reduction reaction (NRR)\cite{34}. Also, the moderate energy gap make it suitable for the application for photocatalysts.

In this work, we employ first-principles computations to design single boron atom anchored g-C$_2$N monolayers (B/g-C$_2$N) and explore their catalytic performance towards N$_2$ reduction. As expected, the decoration of B atom on g-C$_2$N can maintain and enhance slightly the visible light absorption, making the reduction process possibly occur under visible light. In addition, comparing to insufficient ability of g-C$_2$N to adsorb N$_2$, B/g-C$_2$N can bind the N$_2$ strongly, with binding energies of -0.57 eV and -1.40 eV through side-on and end-on patterns, respectively. Surprisingly, the adsorbed N$_2$ can be further reduced into NH$_3$ with a rather low onset potential of 0.07 V and free energy barrier of 0.50 eV through the enzymatic mechanism. The stability of B/g-C$_2$N is systematically evaluated and shows that the designed catalyst is very promising to be synthesized.

\section{Calculation details}
The spin-polarized DFT calculations are performed by using Vienna Ab initio Simulation Package (VASP)\cite{35,36} with projector-augmented-wave (PAW) pseudopotentials\cite{37}. The generalized gradient approximation (GGA) in the form of the Perdew-Burke-Ernzerhof (PBE)\cite{38,39} is used for treating the electronic exchange-correction interactions and a cutoff energy of 500 eV for the plane-wave expansion is adopted. The convergence threshold is 10$^{-5}$ eV and 0.01 eV/{\AA} for energy and force, respectively. A vacuum space is larger than 20 {\AA}, which is set to avoid interactions between two periodic units. The single-atom catalyst are modeled by depositing one B atom on 2 $\times$ 2 $\times$ 1 supercell g-C$_2$N, Brillouin zones are sampled by a Monkhorst-Pack k-point mesh with a 3 $\times$ 3 $\times$ 1 k-point grid. In order to better describe the long-range van der Waals interaction, the DFT-D2 correction\cite{40} proposed by Grimme and co-workers is applied. Hybrid functionals based on the Heyd-Scuseria-Ernzerhof (HSE06) method\cite{41} are employed to obtain the exact band structures and optical absorption spectra. The climbing nudged elastic band (CI-NEB) method\cite{42} as implemented in VTST tools is carried out to search the transition states and saddle points along the reaction pathway, and the activation barriers for electrochemical reaction are calculated using a transferable method reported by Janik et al.\cite{43} Ab initio molecular dynamics simulation (AIMD) by Forcite code\cite{44} with isobaric-isothermal (NPT) ensemble is employed to evaluate the thermodynamic stability of the materials, the simulation is run under 1000 K for 100 ps with time step of 2 fs.

The calculations of Gibbs free energy change ($\triangle$G) for each elemental step are based on the computational hydrogen electrode model proposed by N{\o}rskov et al.,\cite{45,46} which can be determined as follows:
\begin{equation}
\begin{split}
{\Delta}G = {\Delta}E + {\Delta}E_{ZPE} - T{\Delta}S + {\Delta}G_U + {\Delta}G_{pH}
\end{split}
\end{equation}
where $\triangle$E is the difference between the adsorption energies of a given group; $\triangle$E$_{ZPE}$ and $\triangle$S are the changes in zero point energies and entropy, severally, T is the temperature, which is set to be 298.15 K in this work, $\triangle$G$_U$ is the contributions of electrode potential to shift the free energy $\triangle$G at the applied electrode potential ($U$); $\triangle$G$_{pH}$ is the free energy correction of $pH$ ($\triangle$G$_{pH}$ = 2.303 $\times$ k$_B$T $\times$ $pH$ $\approx$ 0.059 $\times$ $pH$), the $pH$ value is set to be zero in this work. The E$_{ZPE}$ and TS for each reaction intermediates can be calculated from the vibrational frequencies, via the following equation:
\begin{equation}
\begin{split}
E_{ZPE} =  \frac{1}{2}\sum_{i=1}^nh\nu_i
\end{split}
\end{equation}
\begin{equation}
\begin{split}
TS = TR\sum_{i=1}^n\{\frac{hc\tilde{\nu}_i}{k_bT(e^{\frac{hc\tilde{\nu}_i}{k_bT}-1})} - In(1-e^{-\frac{hc\tilde{\nu}_i}{k_bT}})\}
\end{split}
\end{equation}
where $R$, $h$, $c$, $k_b$, $\nu$ and $\tilde{\nu}$ are gas constant, Planck constant, light speed, Boltzmann constant, vibrational frequencies and wave number, respectively. Moreover, only the calculation of zero point energy and entropy of reaction intermediates are needed and the contribution of substrate can be offset. The entropies and zero-point energies of molecules in the gas phase are obtained from the NIST database.[http://cccbdb.nist.gov/]

\section{Results and Discussion}
\subsection{B-doped g-C$_2$N}
\begin{figure}[!htb]
\begin{center}
\includegraphics[width=3.4in,angle=0]{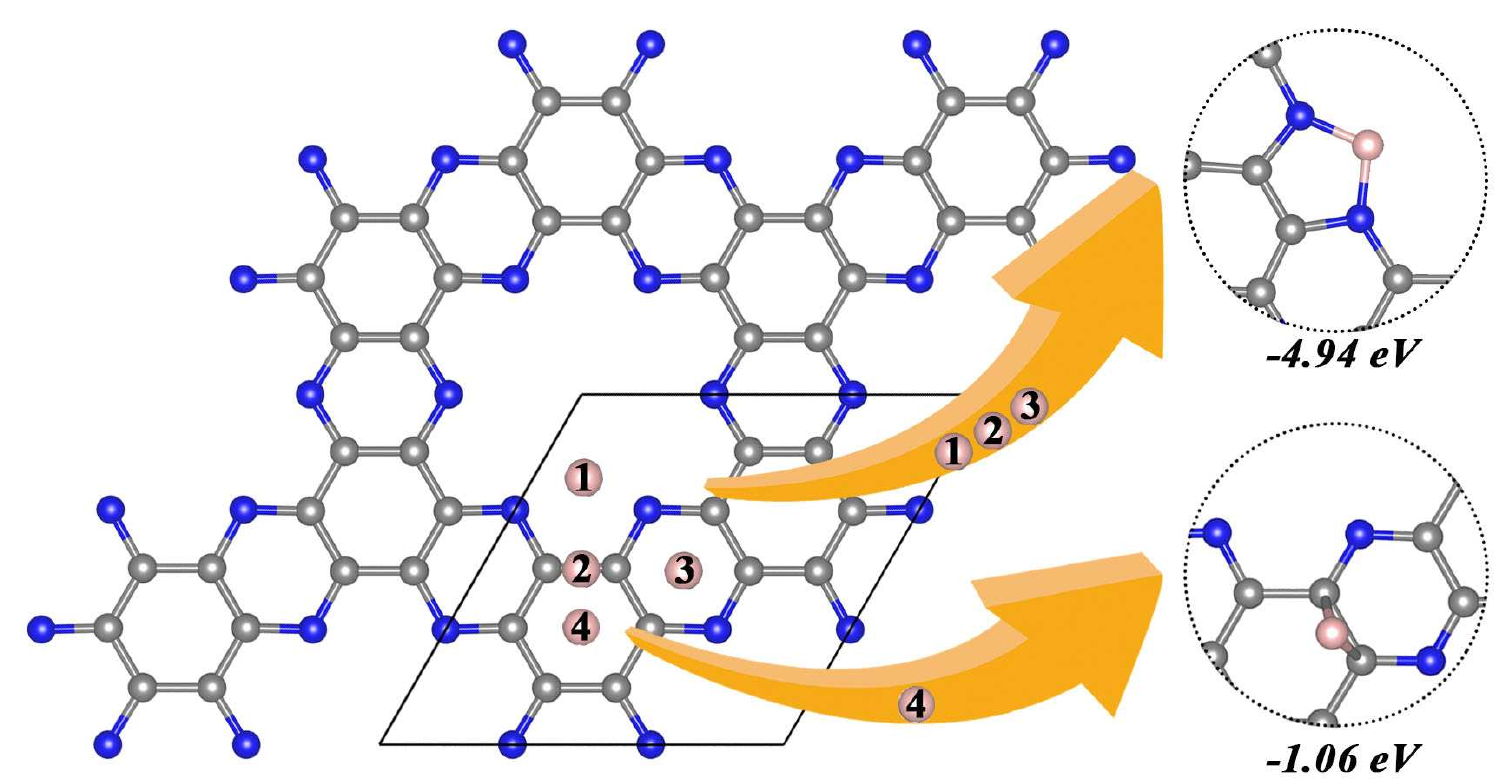}
\caption{\label{fig1}  Initial structures of B atom adsorbed on g-C$_2$N at different sites (at the left panel), and their optimized structures are illustrated on the right panel. The gray, lightpink, and blue balls represent C, B and N atoms, respectively.}
\end{center}
\end{figure}

As shown in Figure 1, on account of the larger radius of B atom than that of C and N atoms, there are 4 rational types of adsorption site for the adsorption of B atom in g-C$_2$N: type-1 is located at the linkage site of two nitrogen atoms, type-2 is positioned in the bridge site of C atom, type-3 corresponds to the middle of six-membered ring which including two N atoms and four C atoms, and type-4 is the middle of six-membered ring of carbon atoms. B atom is added in each kind of adsorption sites. After optimization of these adsorption structures, the type-2 and type-3 transfer to the type-1, and the binding energy is calculated to be -4.94 eV, as for the type-4, B atom moves to the bridge site of C atom and also has negative binding energy of -1.06 eV (Figure 1). Obviously, the binding energy of type-1 is most favorable in energy. High binding energy and large energy difference of the formation energies between the different adsorption types of the B/g-C$_2$N provides significant advantages in terms of experimental synthesis, so the following researches focus on the structure of type-1.

\begin{figure}[!htb]
\begin{center}
\includegraphics[width=3.4in,angle=0]{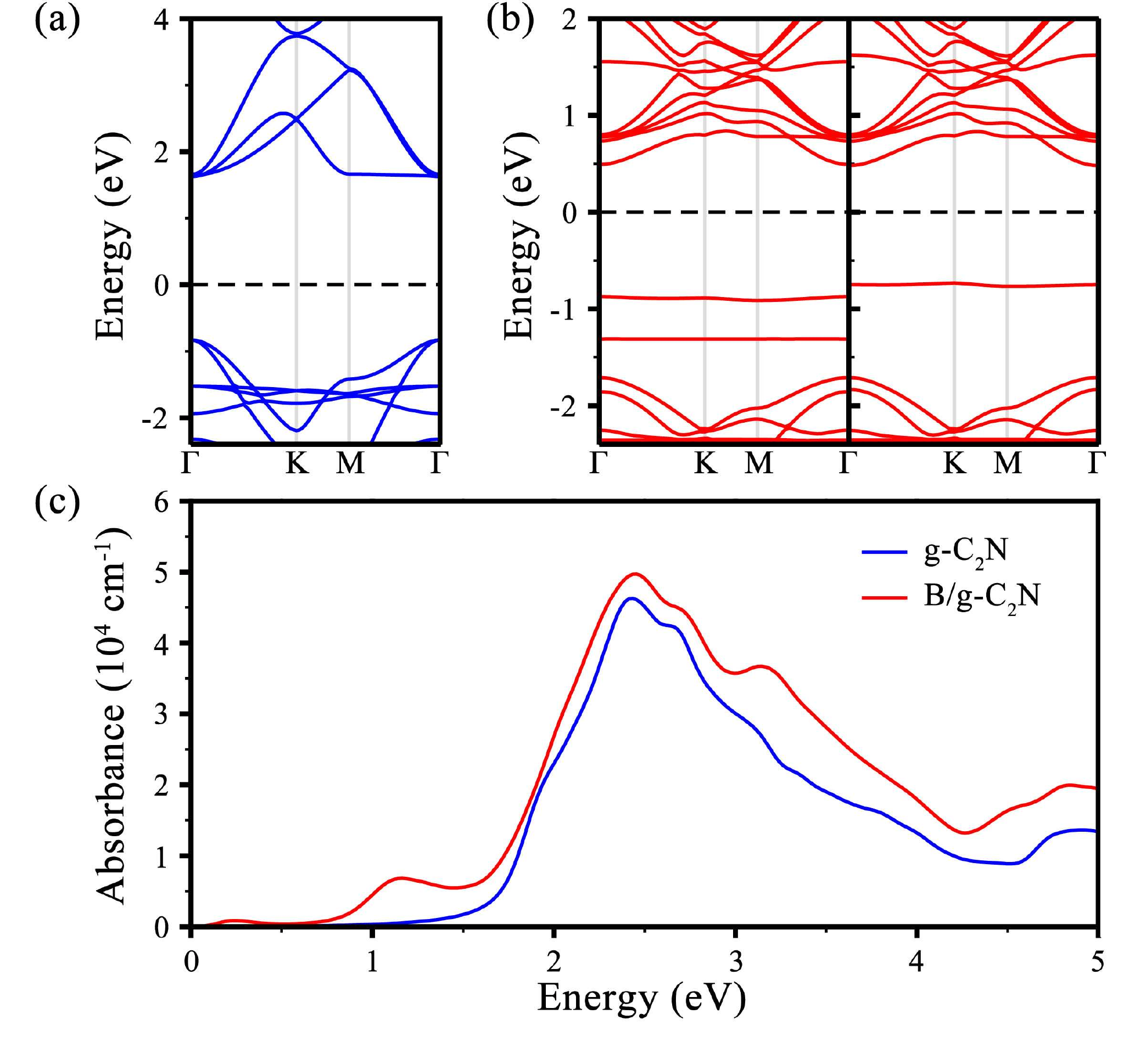}
\caption{\label{fig2} The Band structures of (a) g-C$_2$N and (b) B/g-C$_2$N with spin up (left) and spin down (right). (c) The optical absorption spectra of g-C$_2$N and B/g-C$_2$N. The Fermi level is represented by dash line and set to 0 eV.}
\end{center}
\end{figure}

The deposition of B atom on g-C$_2$N can significantly modify its electronic structure, Figure 2(a)-(c) presents the band structures and optical adsorption spectra of pristine g-C$_2$N and B/g-C$_2$N. The band gap of g-C$_2$N is calculated to be 2.45 eV, which is in accordance with the previous theoretical results\cite{47,48}. With B atom depositing on the g-C$_2$N, introducing a impurity energy level come from B atom lie in the midst of band gap of g-C$_2$N. Therefore, the band gap of B/g-C$_2$N is significant reducing to 1.21 eV, and the non-magnetic g-C$_2$N will possess the magnetic moment of 1 $\mu$$_B$ when B atom interact with two N atoms of g-C$_2$N (moiety of BN$_2$). In order to achieve high-efficiency application in photocatalytic materials, wide and strong optical absorption in the optimal visible-light regions is naturally expected. Thus, we further investigate the optical absorption spectra for the g-C$_2$N and B/g-C$_2$N monolayers. As shown in Figure 2(c), it is noted that the introduction of B atom is found to be further extended the absorption edge to the infrared light region and enhance the visiblelight absorption efficiency slightly, indicating that visible-light utilization of the B/g-C$_2$N is a little more efficient than that of the pure g-C$_2$N monolayer.

\subsection{N$_2$ molecule adsorbed on B/g-C$_2$N}

\begin{figure}[!htb]
\begin{center}
\includegraphics[width=3.2in,angle=0]{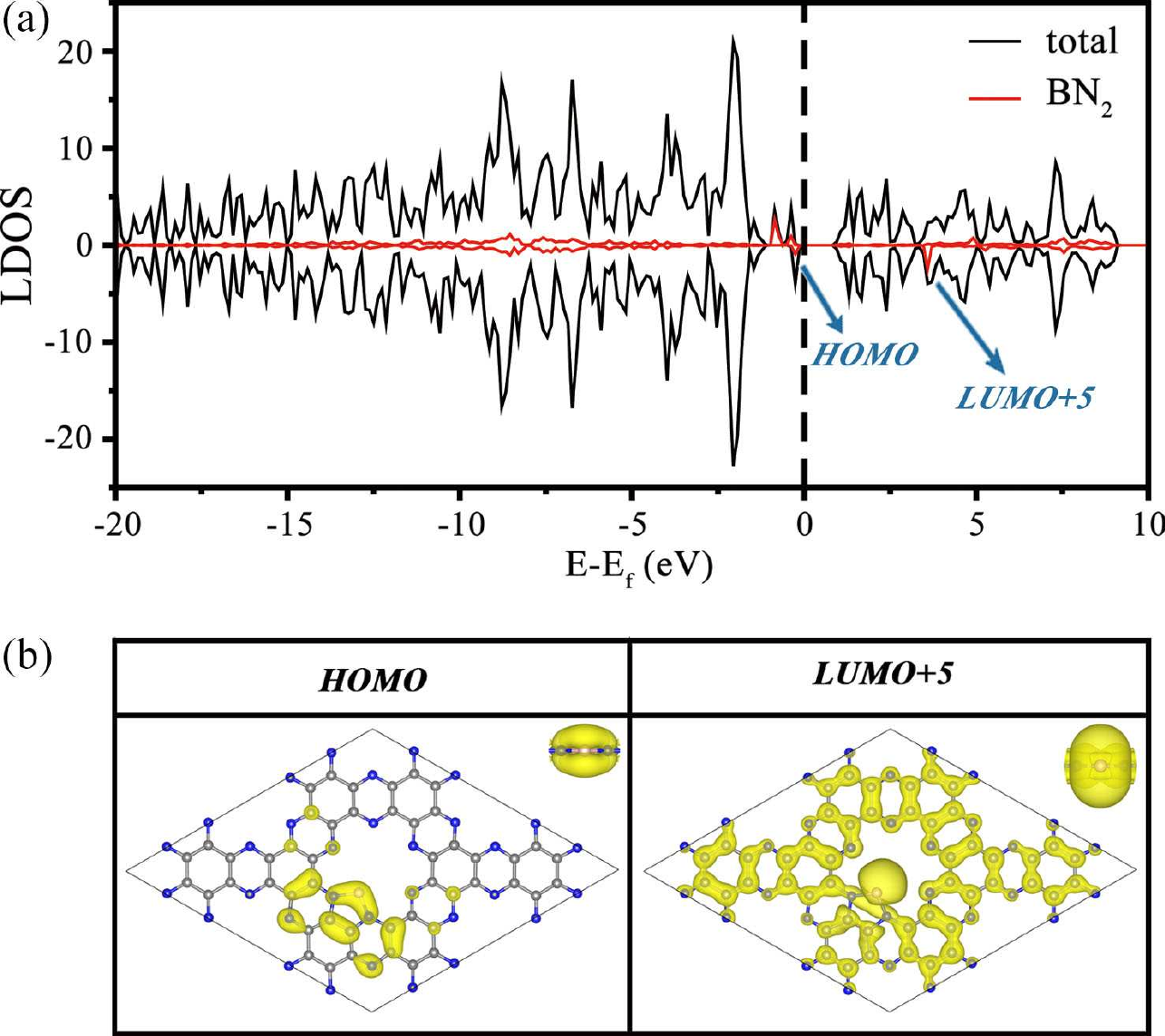}
\caption{\label{fig1} (a) Spin-polarized local density of state (LDOS) of B/g-C$_2$N. (b) The highest occupied molecular orbital (HOMO) and sixth-lowest unoccupied molecular orbital (LUMO+5) of B/g-C$_2$N, where the isosurface value is set to be 0.001 e/{\AA}$^3$. The gray, lightpink, and blue balls represent C, B and N atoms, respectively. The insert shows the orbital of moiety of BN$_2$ on side view.}
\end{center}
\end{figure}

\begin{figure}[!htb]
\begin{center}
\includegraphics[width=3.1in,angle=0]{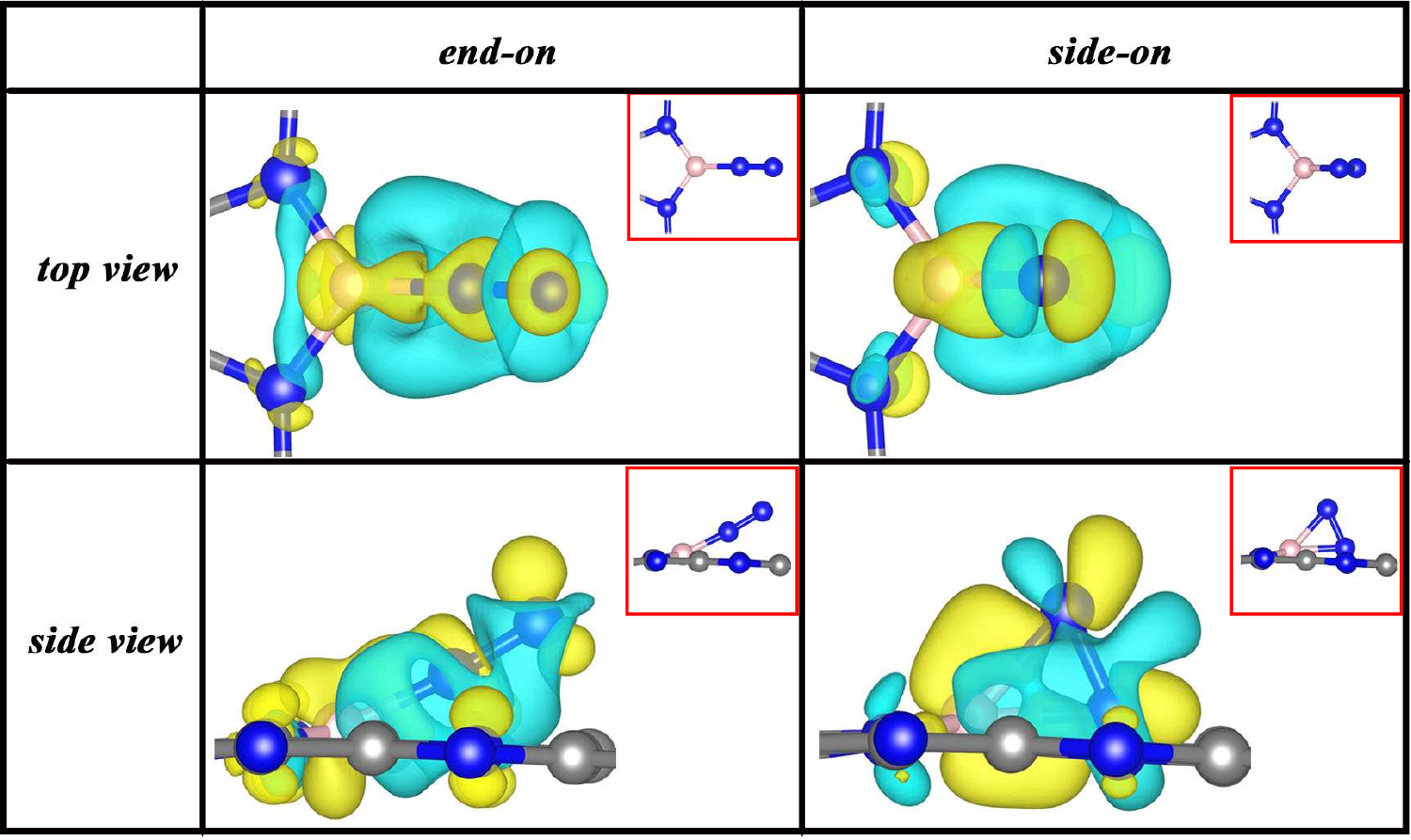}
\caption{\label{fig1} Difference charge density of N$_2$ adsorbed on B/g-C$_2$N, where the isosurface value is set to be 0.02 e/{\AA}$^3$.
The charge accumulation and depletion are shown in yellow and cyan, respectively. The gray, lightpink, and blue balls represent C, B and N atoms, respectively.}
\end{center}
\end{figure}

\begin{table*}[!htb]
\caption{ Adsorption energy (E$_{ads}^{N_2}$, in eV) and bond length (d$_{N-N}$, in {\AA}) of N$_2$ adsorbed on B/g-C$_2$N; bader charge transfer ($\Delta\rho$) from B/g-C$_2$N to N$_2$; Minimum Energy Reaction Pathway (MERP); Rate-Limiting Step (RLS) and onset potential (U, in V) for the electrocatalytic reaction of N$_2$ reduction to NH$_3$ on B/g-C$_2$N.}
\begin{center}
\renewcommand\tabcolsep{12pt}
\begin{tabular} {lllllll} \hline\hline
Adsorption pattern&E$_{ads}^{N_2}$ ($\Delta$G$_{N_2}$)&d$_{N-N}$ ($\triangle$$\rho$)&MERP&RLS&U\\\hline
end-on  &  -1.40 (-0.82) & 1.13 (0.39) &pathway V &  $^*$N-NH$_2$ $\rightarrow$ $^*$N             & 0.74 \\\hline
side-on &  -0.57 (-0.02) & 1.20 (0.73) &pathway V &  $^*$N-$^*$NH $\rightarrow$ $^*$N-$^*$NH$_2$  & 0.07 \\\hline\hline
\end{tabular}
\end{center}
\end{table*}

\begin{figure}[!htb]
\begin{center}
\includegraphics[width=3.4in,angle=0]{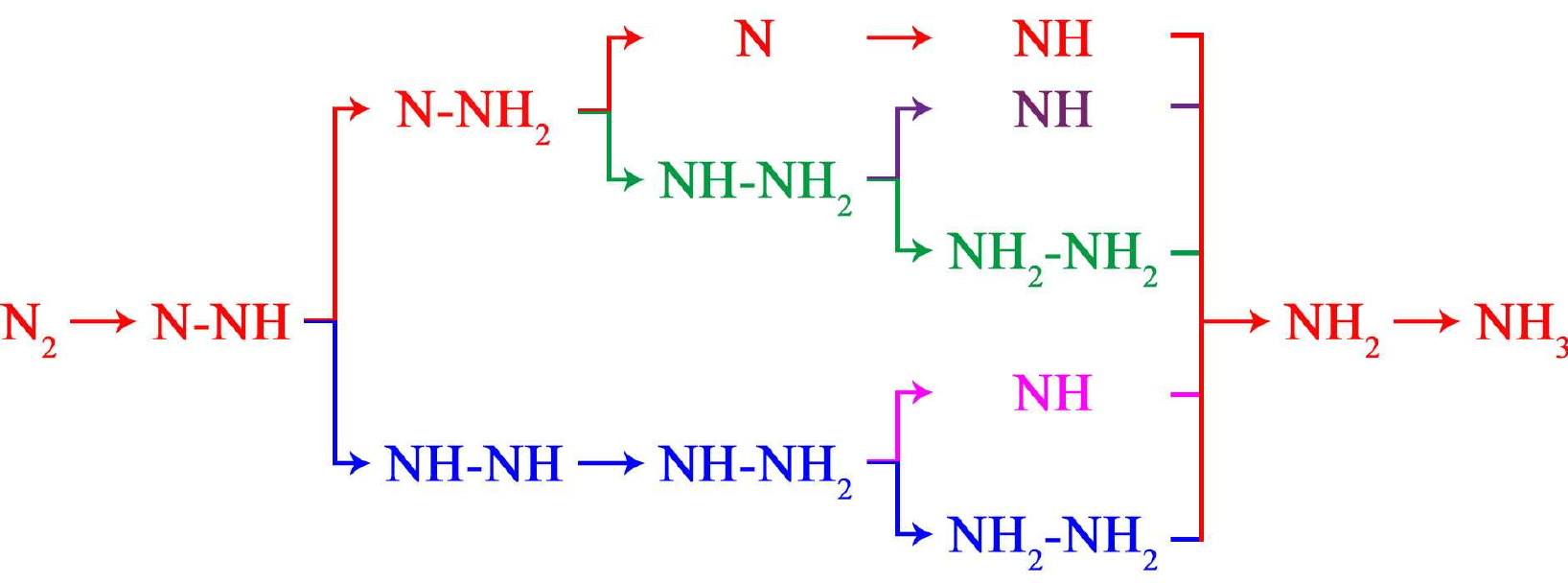}
\caption{\label{fig1} Proposed schematic elementary reactions for N$_2$ reduction to NH$_3$ on B/g-C$_2$N. The red, purple, green, magenta and blue lines represent the reaction pathways I, II, III, IV and V, respectively.}
\end{center}
\end{figure}

As we all known, N$_2$ is adsorbed on catalysts via side-on and end-on adsorption patterns. Therefore, the interaction between N$_2$ and B/g-C$_2$N is investigated. For N$_2$ adsorbed on the B/g-C$_2$N, N$_2$ can only adsorb to B atom of the moiety of BN$_2$ with the release of energy, and the occupied orbital and unoccupied orbital of BN$_2$ near the E$_f$ lie in the highest occupied molecular orbital (HOMO) and sixth-lowest unoccupied molecular orbital (LUMO+5) of B/g-C$_2$N (Figure 3 and S1). And the difference charge density of N$_2$ adsorbed on B/g-C$_2$N are ploted in Figure 4. Adsorption energy (E$_{ads}^{N_2}$, in eV) and bond Length (d$_{N-N}$, in {\AA}) of N$_2$ adsorbed on B/g-C$_2$N; bader charge transfer ($\Delta\rho$) from B/g-C$_2$N to N$_2$; Minimum Energy Reaction Pathway (MERP); Rate-Limiting Step (RLS) and onset potential (U, in V) for the electrocatalytic reaction of N$_2$ reduction to NH$_3$ on B/g-C$_2$N are also listed in Table 1. The shapes of HOMO and LUMO+5 of the moiety of BN$_2$ are similar to $\pi$ orbital (Figure 3), indicating that it has possibility to adsorb N$_2$ via side-on adsorption. This is because there are the better hybridization between N$_2$ and moiety of BN$_2$. As expected, the binding strength between N$_2$ and B/g-C$_2$N is exothermic process with the adsorption free energy of -0.82 and -0.02 eV for end-on and side-on adsorption, respectively. Due to N$_2$ molecule has a lone pair of electrons and the moiety of BN$_2$ has partially occupied $\pi$ orbitals, it can be expected that the interaction between N$_2$ and BN$_2$ is ``acceptance-donation" mechanism. As shown in Figure 4, the charge density difference of B/g-C$_2$N with the adsorption of N$_2$ is calculated to clarify the interaction between B/g-C$_2$N and N$_2$. Interestingly, the ``acceptance-donation" process is found here, which charge accumulation and depletion can be observed for both N$_2$ molecule and B/g-C$_2$N, indicating the N$_2$ can not only accept electrons from B/g-C$_2$N, but also donate electrons to B/g-C$_2$N. Obviously, comparing to donate electrons, N$_2$ tend to accept electrons from B/g-C$_2$N (Figure 4). The bader charge analysis also shows that the N$_2$ gains 0.39 and 0.73 $\mid$e$\mid$ from B/g-C$_2$N for end-on and side-on configurations, respectively (Table 1). Therefore, N$_2$ with side-on adsorption configuration can accept more electrons than end-on adsorption configuration, making the bond length of N$_2$ become longer (1.20 {\AA} v.s 1.13 {\AA}), the increase of N$\equiv$N bond length implies the activation of the inert N$\equiv$N triple bond. We note that the better orbital hybridization can be achieved when N$_2$ is adsorbed onto B/g-C$_2$N with side-on pattern. Therefore, the energy released by side-on adsorption pattern should be more than that released by end-on adsorption pattern. However, due to the ``acceptance-donation" interaction mechanism, more electrons are obtained by 1$\pi$$^*$ anti-bonding orbital of N$_2$ for side-on adsorption pattern, as mentioned in our previous work\cite{49}, which will lead to the longer N$\equiv$N bond length. Therefore, the energy released by side-on pattern is less than that released by end-on pattern, because most of the energy is used to activate the N$\equiv$N triple bond. As expected, as shown in Table 1, the adsorption of N$_2$ to B/g-C$_2$N is an exothermic process with the adsorption free energy of -0.82 eV and -0.02 eV for end-on and side-on adsorption patterns, respectively.

\subsection{Electrocatalytic NRR}

\begin{figure}[!htb]
\begin{center}
\includegraphics[width=3.2in,angle=0]{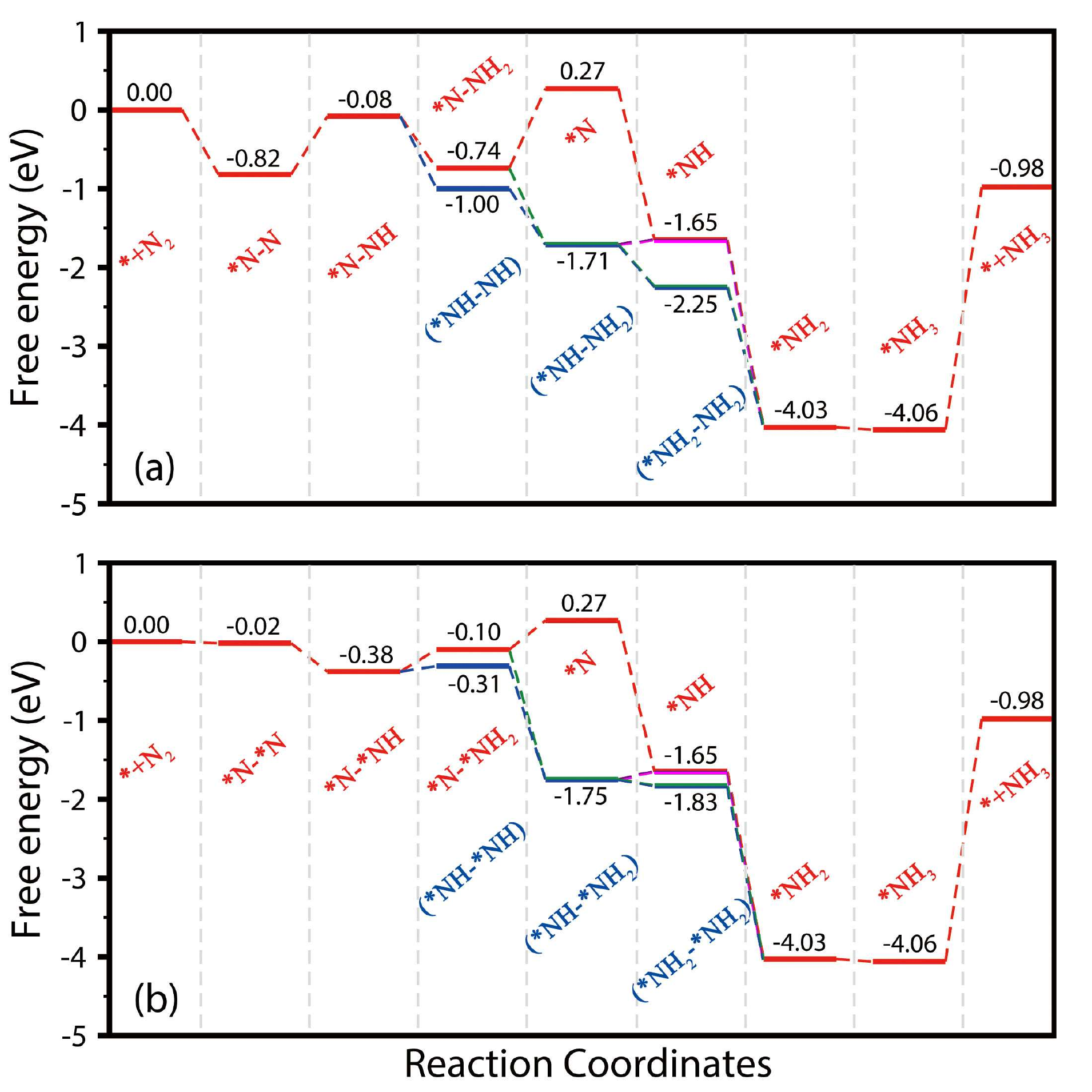}
\caption{\label{fig1} Gibbs free-energy change ($\Delta$G) for N$_2$ reduction to NH$_3$ on B/g-C$_2$N through (a) end-on and (b) side-on pattern. ``*" denotes the catalyst (B/g-C$_2$N). The red, purple, green, magenta and blue lines represent the reaction pathways I, II, III, IV and V, respectively, as shown in Figure 5. The energy of the entrance, including the isolated B/g-C$_2$N and free N$_2$ molecule, is set to zero as the reference.}
\end{center}
\end{figure}

\begin{figure*}[!htb]
\begin{center}
\includegraphics[width=6in,angle=0]{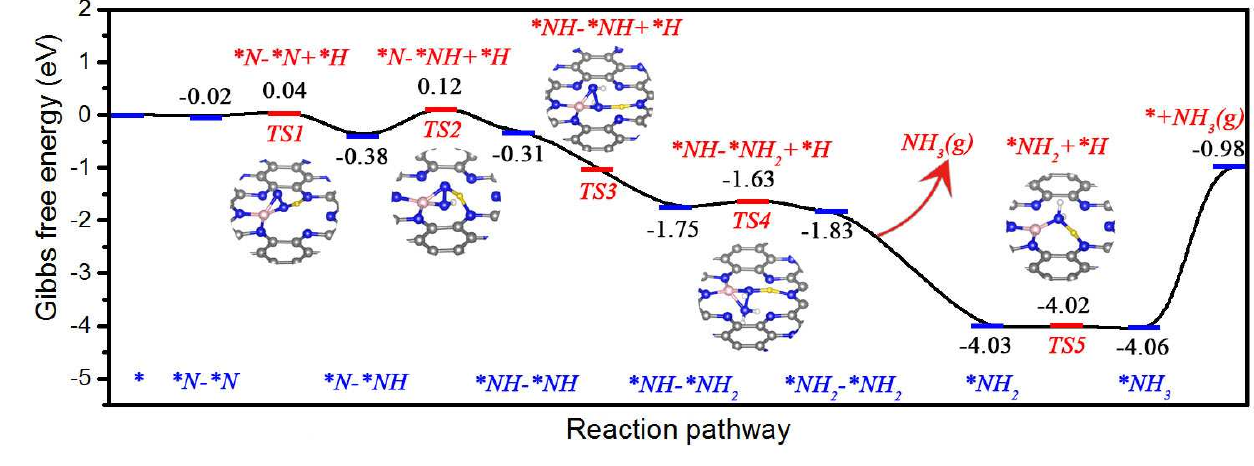}
\caption{\label{fig1} Minimum energy pathway for N$_2$ reduction into NH$_3$ catalysed by B/g-C$_2$N through side-on pattern. The energy of the entrance, including the isolated B/g-C$_2$N and free N$_2$ molecule, is set to zero as the reference. The structures of the Transition States (TS) are indicated. The gray, lightpink, blue and white balls represent C, B, N and H atoms, respectively. The reactive hydrogen atom during the N$_2$ electro-reduction is highlighted in golden in each step.}
\end{center}
\end{figure*}
Generally, as shown in Figure 5, the possible reaction pathways of N$_2$ reduction to NH$_3$ on B/g-C$_2$N will follow five possible reaction pathways whether the adsorption configuration is end-on or side-on adsorption pattern. The assessment of the catalytic performance about B/g-C$_2$N for the reduction of N$_2$ into NH$_3$ by calculating NRR is evaluated by calculating the Gibbs free energy change ($\triangle$G) in every hydrogenation steps, and the reaction process involve six net proton coupled electron transfer (PCET) steps. Figure 6 displays all of the reaction pathways with the corresponding Gibbs free energy profile for B/g-C$_2$N catalyze N$_2$ reduction. The initial adsorption of N$_2$ molecule for the end-on and side-on adsorption patterns release free energy of 0.82 eV and 0.02 eV, respectively. For the end-on adsorption pattern, there is a maximum $\triangle$G ($\triangle$G$_{max}$) of 1.01 eV in the third PCET step (*NNH$_2$ + H$^+$/e$^-$ $\rightarrow$ *N + NH$_3$) of the reaction pathway I. At the same time, as shown in Figure {\color{blue}6}{\color{red} \sout{5}}(a), a relatively small Gibbs free energy barrier of 0.06 eV is found in the fourth PCET step (*NH-NH$_2$ + H$^+$/e$^-$ $\rightarrow$ *NH + NH$_3$) of the reaction pathway IV, and there is no Gibbs free energy barrier for the rest of hydrogenation steps. We also notice that, in the end-on pattern, all reaction pathways need to cross a larger reaction barrier (0.74 eV, the first protonation: *N$_2$ + H$^+$/e$^-$ $\rightarrow$ *NNH), which means that it is very difficult for nitrogen reduction into ammonia at ambient conditions.

As for the five possible reaction pathways of the side-on adsorption configuration, as shown in Figure 6(b), the $\triangle$G$_{max}$ for the reaction pathway I is 0.37 eV in the third PCET step (*N-*NH$_2$ + H$^+$/e$^-$ $\rightarrow$ *N + NH$_3$), which is the largest $\triangle$G$_{max}$ among these five reaction pathways. Meanwhile, the $\triangle$G$_{max}$ for the reaction pathway V is 0.07 eV in the second PCET step (*N-*NH + H$^+$/e$^-$ $\rightarrow$ *NH-*NH), which is the smallest $\triangle$G$_{max}$ among these five reaction pathways. It is worth noting that, in this reaction pathway V, every elementary steps are exothermic except for the second protonation to form *NH-*NH. So we come to a conclusion that, for the side-on adsorption pattern, the onset potential only require 0.07 V to eliminate the increase in Gibbs free energy, and the NRR reaction can be carried out smoothly. Such a smaller onset potential means that it is very easy for nitrogen reduction into ammonia at ambient conditions.

\begin{figure*}[!htb]
\begin{center}
\includegraphics[width=6in,angle=0]{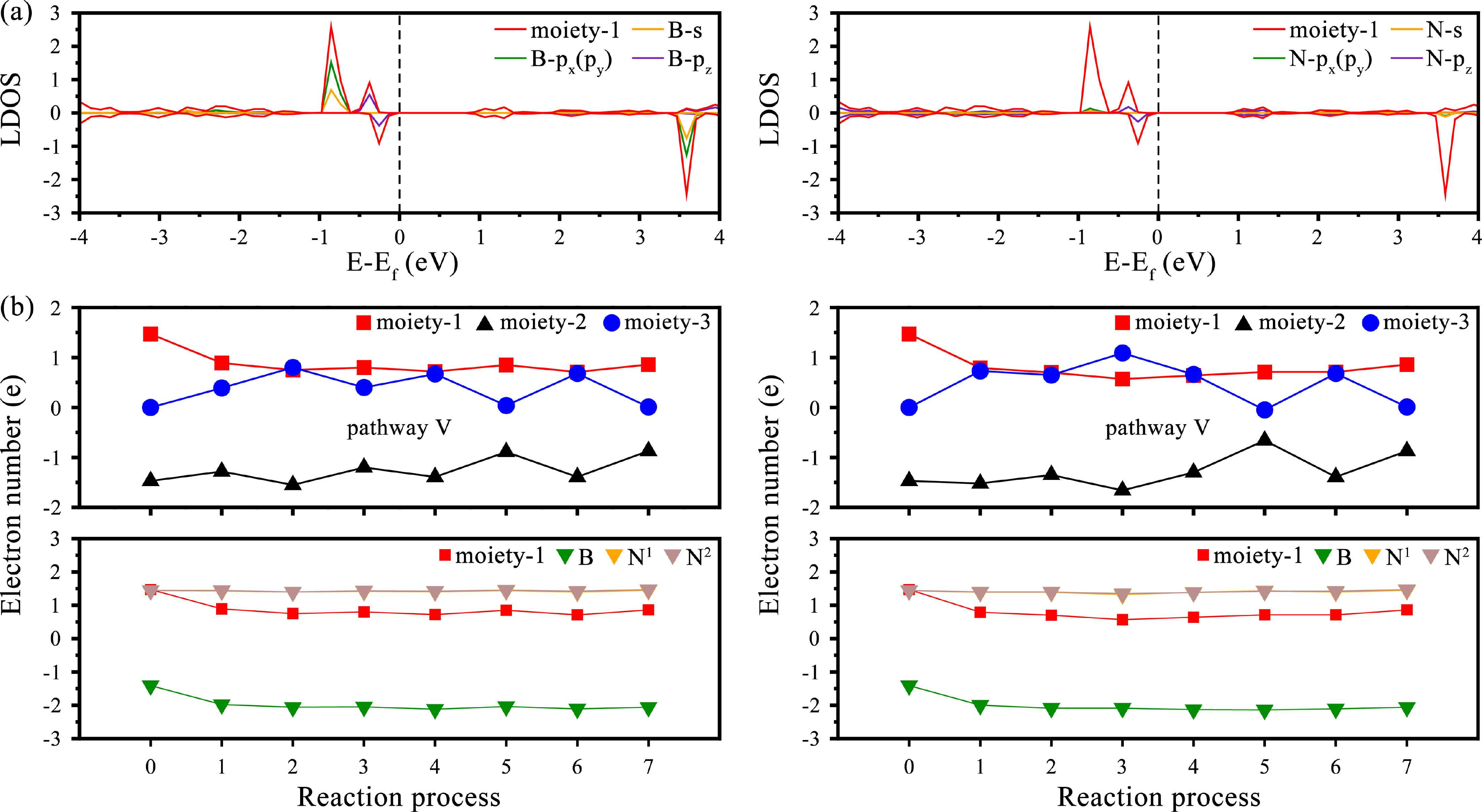}
\caption{\label{fig1} (a) Spin-polarized local density of state (LDOS) of the moiety of BN$_2$. (b) Charge variation of the three moieties along the MERP (reaction pathway V) of N$_2$ with end-on (left) and side-on (right) adsorption configuration. The moiety-1, moiety-2 and moiety-3 represent BN$_2$ of B/g-C$_2$N, other parts of B/g-C$_2$N and adsorbed N$_2$, respectively. N$^1$ and N$^2$ are the N atom which bond to B atom. Reaction process 0, 1, 2, 3, 4, 5, 6, 7 represent the isolated state, N$_2$ adsorption reaction, initial, second, third, fourth, fifth and sixth protonation reaction, respectively.}
\end{center}
\end{figure*}

Furthermore, in order to enhance the understanding of the reaction mechanism, it is also vital to find out the transition states in the reaction pathway. Figure 7 shows the structures and energies of the Transition States (TS) in the minimum energy pathway over the reaction pathway V with side-on pattern. It's easy to find that the maximum free energy barrier of 0.50 eV lies in the second PCET processes (*N-*NH + H$^+$/e$^-$ $\rightarrow$ *NH-*NH), i.e. TS2. As for the rest of the reaction processes, the free energy barriers are all lower than that of TS2. The free energy barriers for TS1, TS4 and TS5 are only 0.06 eV, 0.12 eV and 0.01 eV, respectively. At the same time, there is no barrier for TS3. Considering that the previous reactions have released energy of 0.38 eV, this step (*N-*NH + H$^+$/e$^-$ $\rightarrow$ *NH-*NH) only requires 0.12 eV. That is to say, the extra energy input for the NRR reaction based on the reaction pathway V with side-on pattern only needs 0.12 eV.

Interestingly, as shown in Figure 2, the introduction of B atom into g-C$_2$N is found to be further extended the absorption edge to the infrared light region and enhance the visiblelight absorption efficiency slightly. The maximum absorption peak is located in 2.45 eV, indicating B/g-C$_2$ possess outstanding ability to absorb certain energies of visible light. And the smaller peak located at 1.20 eV originate from the doping of B atom. Due to the fact that the maximum free energy barrier is only 0.50 eV, photon energy absorbed by the B/g-C$_2$N is enough to overcome this barrier, highlighting that the outstanding catalytic performance of B/g-C$_2$N for NRR.

 \begin{figure}[!htb]
\begin{center}
\includegraphics[width=3in,angle=0]{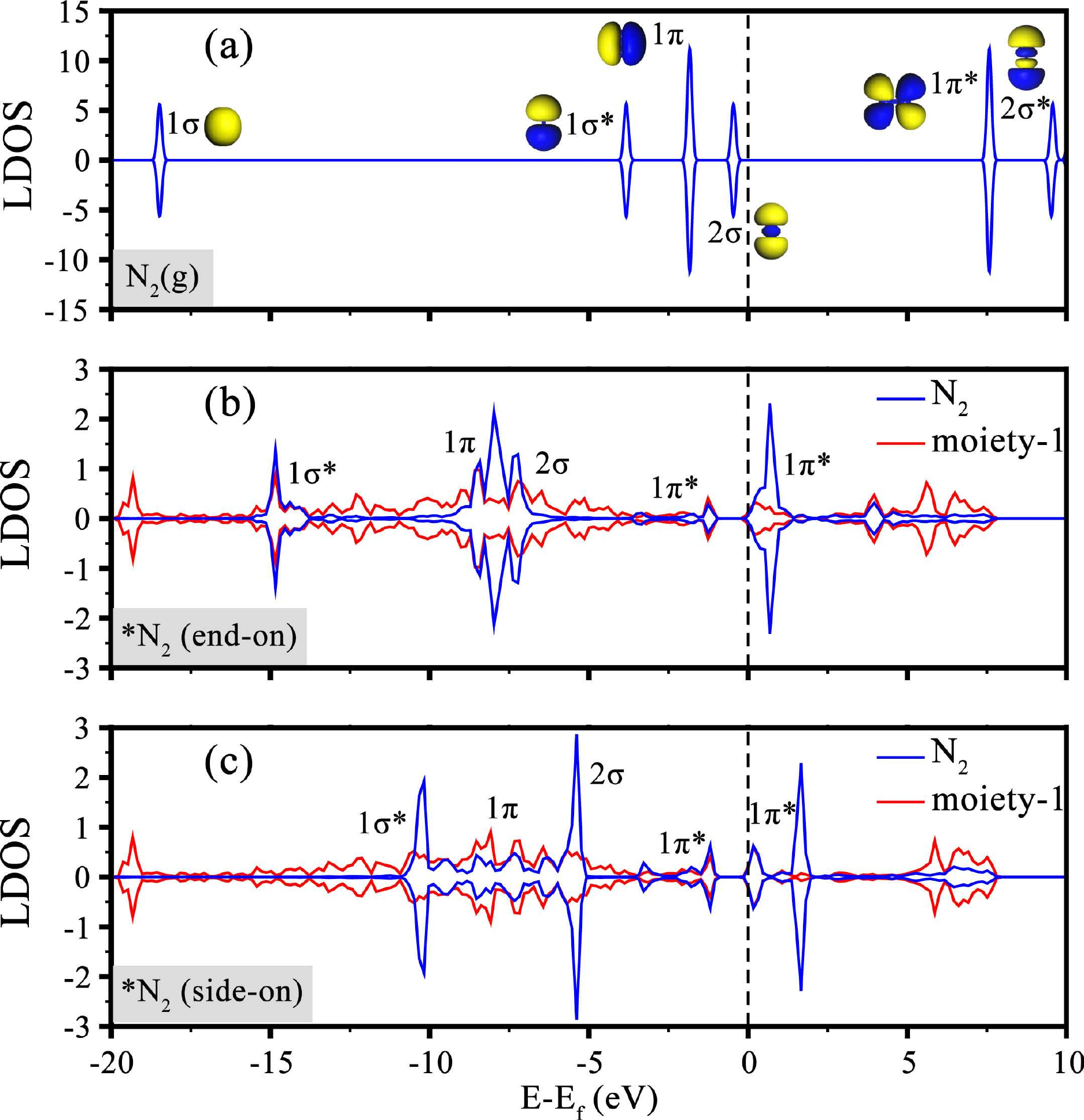}
\caption{\label{fig1} Spin-polarized local density of state (LDOS) of (a) free N$_2$ molecule, N$_2$ adsorbed on B/g-C$_2$N with (b) end-on pattern and (c) side-on pattern. The blue and red lines represent the total LDOS of N$_2$ and BN$_2$ of B/g-C$_2$N (moiety-1), respectively. The Fermi level is set to 0 eV.}
\end{center}
\end{figure}

As seen from the figure 8 and S2, the bader charge analysis of whole reaction show that the electrons of transfer to N$_2$ are mainly come from the moiety of BN$_2$ of B/g-C$_2$N (moiety-1) when N$_2$ in initial adsorption whether end-on adsorption or side-on adsorption, indicating the BN$_2$ play the key role to capture and active N$_2$. Thereinto the B atom transfer more electrons to N$_2$ than the N atom of BN$_2$, because the occupied orbital (HOMO) and unoccupied orbital (LUMO+5) of BN$_2$ consist largely of s, p$_x$(p$_y$) and p$_z$ orbitals of B atom, besides, N$_2$ direct contact with B atom (figure 8a). However, in the course of the reaction, the BN$_2$ work as a transportation center to move the electrons of other parts of B/g-C$_2$N (moiety-2) which act as a electron reservoir to the N$_x$H$_y$ (moiety-3), because the amounts of electrons for moiety-1 almost no change during the whole reaction. The introduction of B atom make g-C$_2$N can not only attract N$_2$ appropriately, but also carry the electrons of reservoir to the N$_2$ effectively.

In order to find out the origin of activity of B/g-C$_2$N, the local density of states (LDOS) of free N$_2$ molecule and B/g-C$_2$N with adsorbed N$_2$ molecule are calculated and shown in Figure 9. The molecular orbitals of free N$_2$ molecule mainly include 2$\sigma$, 2$\sigma$$^*$, 1$\pi$, 3$\sigma$, 1$\pi$$^*$, and 3$\sigma$$^*$ molecular orbitals. When N$_2$ is adsorbed on B/g-C$_2$N, the molecular orbitals of N$_2$ will hybridize with the HOMO and LUMO of moiety-1 of B/g-C$_2$N intensively, as shown in figure 9b-c. For the end-on adsorption pattern, the 1$\pi$$^*$ orbital of N$_2$ can accept the electrons form the HOMO of moiety-1 of B/g-C$_2$N, leading the 1$\pi$$^*$ orbital is partially occupied, and the 2$\sigma$$^*$, 1$\pi$, 3$\sigma$ orbitals can donate the electrons to the LUMO of moiety-1 of B/g-C$_2$N results in the decreases of their strength, further confirm the ``acceptance-donation" interaction mechanism, the Bader charge analysis show that N$_2$ will get 0.39 $\mid$e$\mid$ net charge from B/g-C$_2$N. As for side-on adsorption pattern, the 2$\sigma$$^*$, 1$\pi$, 3$\sigma$ orbitals have also changed and their strength also decreases, the difference is that the degree of the ionization of 1$\pi$ orbital and the filling of 1$\pi$$^*$ orbital are stronger, according with our previous research\cite{49}. From the Bader charge analysis, we can find that N$_2$ will get 0.73 $\mid$e$\mid$ net charge from B/g-C$_2$N, which means the 1$\pi$$^*$ orbital of N$_2$ can accept more electrons form the HOMO of moiety-1 of B/g-C$_2$N in side-on adsorption pattern than end-on adsorption pattern, that's why the difference of d$_{N-N}$, the longer d$_{N-N}$ means the more active of N$_2$. So, the catalytic performance of B/g-C$_2$N with the side-on pattern of N$_2$ is the best reaction pathway.

Moreover, the stability of catalyst is also very important to the application in realistic case. We have examined the stability by using AIMD simulations. As shown in Figure S3, when temperature increases to 1000 K, the structure of B/g-C$_2$N is a slight of distortion in vertical plane, but a horizontal surface is still no great changes, showing the outstanding thermostability.

\section{Conclusion}
In summary, we have demonstrated that B/g-C$_2$N can be a excellent photocatalyst for N$_2$ reduction to NH$_3$ by performing the first-principles calculations. The calculated results indicate that the introduction of B atom to g-C$_2$N can not only maintain and improve slightly the optical absorption in the visible-light region relative to the pure g-C$_2$N, but also act as capture and activate center of N$_2$ through ``acceptance-donation" mechanism. The low thermodynamics reaction free energy of 0.07 eV and free energy barrier of 0.50 eV, far below the photon energy absorbed by the B/g-C$_2$N, making B/g-C$_2$N as photocatalyst holds great potential for the reduction for the N$_2$ fixation. In addition, the strong binding energy offer high probability in terms of experimental synthesis. Furthermore, B/g-C$_2$N shows high thermal stability even though the temperature up to 1000 K.

\begin{acknowledgments}
This research is supported by the Natural Science Foundation of China (11375091), the Natural Science Foundation of Zhejiang (LY18A040003), the Natural Science Foundation of Ningbo (2018A610220), the K.C. Wong Magna Fund in Ningbo University. The computation is performed in the Supercomputer Center of NBU.
\end{acknowledgments}

\end{document}